# Ab Initio Physics Calculations for Borophene for Electronic Devices

Clifford M. Krowne
Electromagnetics Technology Branch, Electronics Science & Technology Division,
Naval Research Laboratory, Washington, DC 20375

and

Xianwei Sha*

General Dynamics Information Technology Corporation, Falls Church, VA 22042 and Information Technology Division, Center for Computational Science, Naval Research Laboratory, Washington, DC; *Onsite Contractor

## Abstract

Moving beyond traditional 2D materials is now desirable to have switching capabilities (e.g., transistors). Here we propose using borophene because, as we will show in this letter, obtaining regions of the electronic bandstructure which act as valence and conduction bands, with an apparent bandgap, may be obtainable in the foreseeable future. Here for particular allotropes of borophene, density of states (DOS) and electronic bandstructure diagrams with $\varepsilon(\mathbf{k})$ vs $\mathbf{k}$ are found with much improved accuracy by ab initio quantum calculations using hybrid functionals of several types. This procedure should allow much better insight into how to obtain acceptable materials.

## Introduction

Just to the left of carbon in the Periodic Table is boron. Creation of borophene, consisting entirely of boron atoms in a 2D sheet form [1], makes one recall ideas of using boron nitride (BN) as a material for electronics. However, the intrinsic simplicity of dealing with a monoatomic material is tremendously attractive. One is reminded of the initial fascination with silicon and germanium in the 1950s and 1960s. Silicon grew into the workhorse of electronics for transistors for digital computers and lower frequency RF electronics. In particular, one can think of the use of silicon for affordable cell phones and digital cameras.

However, for high frequency electronic devices, it is well known that antimonide-based compound semiconductors can have very high mobility and offer important advantages for low-power, low-noise RF electronics [2]. Unfortunately, antimonide-based materials suffer several important disadvantages including In and Sb do not have commercially-viable domestic sources [3]; InSb-based electronics can be difficult to monolithically integrate with other electronics technologies (Si-based CMOS digital, GaN-based high-power RF, *etc.*) because of the large lattice-mismatch between the materials families; there is continuing pressure to move to less toxic materials electronics (e.g. ROHS standard); finally, while III-V compound semiconductors have demonstrated $f_T$ and $f_{Max}$ cutoff frequencies in the low THz regime, even higher cutoff frequencies are needed for future commercial and DoD systems applications and it is not clear that III-V semiconductors can meet those demands [4]. New 2D materials with band gaps are predicted to have clear advantages for these applications, if the parasitic resistances can be reduced sufficiently. For all of these reasons, the time is ripe for investigation of new based 2D electronics materials such as borophene that do not suffer from these limitations.

It is known that borophene has allotropes which are planar, and the predicted electronic properties hold interest for high speed transport of carriers [5] – [9]. Growth and preparation of the material for further exploitation in electronic devices, for example, require capping the exposed boron monolayer, due to its reactivity with air.

Borophene comes in allotropes which are planar and display bandgaps appropriate for semiconductor electronics. The various allotropes of borophene are hexagonal honeycomb lattices, with some proportion of the hexagon interiors filled by an additional boron atom. Bandstructure can be selected to be either Dirac type with its relativistic analogy to low mass particles (and no bandgap), or parabolic type with its more massive characteristic of semiconductor action (and a bandgap). The problem with the use of nanoribbons, it was discovered both experimentally and theoretically in graphene, for example, is that large amounts of edge scattering of the carriers results, negating the advantage of opening up a gap with a finite width monatomic sample [10]. This may be expected to be a problem with the recent borophene research directed toward nanoribbon preparation [11].

Turning our attention to borophene, besides its interest for electronics, this new material could revolutionize sensors, batteries and catalytic chemistry [12]. Borophene has excellent possibilities as anode material for lithium-ion batteries, uses in catalysis, and its ability to detect atoms and molecules [13], [14]. Since H atoms easily stick to its surface, hydrogen storage capabilities exist. Finally, borophene can catalyze the breakdown of $H_2$ into its ions, $H_2O$ into H and O ions, and perform $CO_2$ electroreduction.

Although nanowires in the sense of nanotubes have been studied theoretically and experimentally for electronic device possibilities [15], [16], we avoid that here because tube alignment is known to be a difficult challenge with substantial fabrication costs. Similarly, we avoid two terminal devices [17] – [19] to focus instead on transistor capability, where the largest impact on electronics will occur [20], [21].

Synthesis of 2D sheet boron by first-principle studies in [22], obtaining chemical potential for various vacancy concentrations x = 0, 1/9, 1/3, on substrates Cu(111), Ag(111), Au(111), Mg-terminated $MgB_2$, and Ti-terminated $MgB_2$ surfaces, was done. Top cross-section structures shown for B sheets in vacuum and on Ag(111) and Mg-terminated $MgB_2$ (0001) surfaces. Adhesion energies provided for B sheets to substrates, as well as cross-sectional electron distributions.

## First Principles Calculations

Now look at some results we have obtained. Figure 1 shows atomic structures of the α−1 borophene and β−1 borophene allotropes. Figure 2 gives the density of states (DOS) for these two allotropes of borophene. Figure 3 provides the energy ε(**k**) vs. **k** bandstructure through symmetry points in the Brillouin Zone for these two borophene allotropes. From the literature, it is known that first-principles calculations using the traditional generalized gradient approximation (GGA) as the exchange correlation functional usually predict no band gap for the borophene allotropes. We performed density functional theory (DFT) calculations using Perdew--Burke--Ernzerhof (PBE) GGA functional for both alpha-1 and beta-1 borophene nanostructures, and find no band gap for both structures, in agreement with previous publications. We show the calculated band structure and density of states (DOS) of alpha-1 borophene in an expanded view in Fig. 4.

Wu et al. [6] demonstrated that when they use hybrid functional PEB0, which is supposed to be more accurate than GGA as the exchange-correlation functional, in their DFT calculations, several alpha-types of borophene allotropes showed band gaps of ~ 1 eV (Fig. 3), while other types of allotropes still have no gaps. By borophene allotrope crystallographic preparation, the type of band structure can be selected to be either Dirac type with its relativistic analogy to low mass particles (and no bandgap), or parabolic type with its more massive characteristic of semiconductor action (and a bandgap). In either case the desirable properties should be maintainable over actual 2D monoatomic sheets, and not necessitate the tactic of creating nanoribbons to craft a bandgap as done for graphene, for example.

## Results Using Quantum Ab Initio Approach

Wu et al. [6] used only PEB0 as the hybrid functional in their research on borophene nanostructures. It is important to know how different types of hybrid functional [23] might affect the results, so this article addresses all three types of hybrid functionals, PEB0 [24], HSE [25], and B3LYP ]26] to examine borophene nanostructures, with the technical objectives to evaluate and identify the accuracy and efficiency of recently-developed hybrid functionals and determine effects on the borophene electronic band structure.

We use quantum espresso, an integrated suite of open-source computer codes for electronic-structure calculations and materials modeling at the nanoscale. Quantum espresso is based on density-functional theory, plane waves basis set, and pseudopotentials, with many features to examine the ground-state calculations, including structural optimization, molecular dynamics, potential energy surfaces, electrochemistry and special boundary conditions, response properties, spectroscopic properties, quantum transport, and platforms. Quantum espresso implements all different types of popular hybrid functional: PBE0, B3LYP, HSE  hybrid GGA, and works with both the  normal-conserving pseudopotential (NCPP) and ultrasoft-pseudopotential (USPP) with some limitations. Hybrid functional DFT calculation is the newly developed and evolving feature in quantum espresso, which we exploit here.

In typical DFT calculations, in order to obtain the band structure of a given material, one usually first performs a Self-Consistent Field (SCF) calculation to approach convergence, and then a non-SCF calculation using the SCF charge density but on a much finer k-point mesh to obtain the band energies on enough k-points in the Brillouin zone. Unfortunately, such non-SCF DFT calculations using hybrid functional are still not supported in the latest version of quantum espresso. In order to obtain the band structure from hybrid functional DFT calculations, it is necessary to perform SCF calculations on much finer k point mesh, which are usually very computationally demanding. Since our supercell has relatively large size (~30 angstrom) along the 4direction perpendicular to the borophene plane, we only need one k point along that direction. The k point sets we tested include: 1. 4 x 4 X 1; 2. 6 x 6 X 1; 3. 8 x 8 X 1; 4. 12 x 12 x1; 5. 16 x 16 X 1. Hybrid functional calculations using 16x16x1 k-point mesh turns out to be too computationally demanding, which the tests are not able to approach SCF convergence after running large-scale parallel calculations using hundreds of cores for 120 hours on the supercomputer Thunder at AFRL DSRC. For smaller k-point mesh such as 12x12x1, we are able to approach SCF convergence using hundreds of cores with a reasonable amount of time, namely, about 70 hours. The tests and computational work have been performed on supercomputers Thunder and Mustang at AFRL DSRC, Excalibur at ARL DSRC, and Polar at NRL-DC.

    For alpha-1 borophene nanostructure shown in the left side of Fig. 1, we use 12x12x1 k point mesh in the hybrid functional DFT calculations, which produces totally 49 k-points in the irreducible part of the Brillouin zone. We calculated band structures of alpha-1 borophene using PEB0, HSE and B3LYP as the hybrid functional, respectively. B3LYP predicts a slightly different band structure from PBE0 and HSE. The band structures from PBE0 and HSE are essentially identical, although the exchange-correlation functionals used are quite different, as shown in the quantum espresso output files: Exchange-correlation = PBE0 ( 6 4 8 4 0 0); Exchange-correlation = HSE ( 1 4 12 4 0 0).

    All our hybrid functional calculations show that alpha-1 borophene nanostructure has essentially no band gap, in contrast to Wu et al. [6] who predicted a small band gap using PEB0. Several effects might account for the discrepancies here. First, different pseudopotentials have been used. Second, and the most likely reason, is due to the structural relaxation. We performed relaxations using the structure from Wu et al. [6] as the initial structure, and we notice some changes in the supercell dimensions after the structural relaxation. Borophene is known to have many different metastable allotropes where the energies of different allotropes are pretty close. We also notice that for all our hybrid functional calculations for borophene alpha-1 structure, there is only one band crossing near the Fermi level, and along that band there is only one point (Y point in the Brillouin zone) that has energy below the Fermi level. Slight change in the *a* direction on the borophene plane (corresponding to the Y direction in the Brillouin zone) might change the band energy around the Y point. For detailed comparisons between the hybrid functionals, the Fermi level is shifted to zero and the calculated band structures are plotted together, as shown in Fig. 5. It can be seen that the band structures of PBE0 and HSE are essentially identical for both the valence band and the conduction band. The calculated valence bands from B3LYP are in close agreements with PBE0 and HSE, but display some noticeable discrepancies in the conduction band regime, especially close to the Y and S points in the Brillouin zone.

    The calculated band structures of beta-1 borophene nanostructure using PBE0, HSE and B3LYP as the hybrid functional have also been found. There are totally 74 k points in the irreducible part of the Brillouin zone due to the symmetry of the crystal structure. Similar to the alpha-1 borophene results, the band structures calculated from PBE0 and HSE are essentially identical. The calculated band structure of beta-1 borophene using B3LYP agree well with HSE and PBE0 in the valence band regime, but shows noticeable differences in the conductance band regime, especially around

the Y point in the Brillouin zone. All the hybrid functional calculations show that there is more than one band crossing the Fermi level, indicating the metallic nature of beta-1 borophene, in agreement with previous research.

## Conclusions

We tested and evaluated various forms of hybrid functionals in quantum espresso for borophene nanostructures. For both alpha-1 and beta-1 borophene nanostructures, the calculated band structures from PBE0 and HSE are essentially identical, and the band structure calculated using B3LYP are in close agreements with HSE and PEB0 results in the valence band regime, but show some noticeable differences in the conductance band regime. All the calculated band structures of alpha-1 borophene nanostructure show no band gap, in contrast to Wu et al. [6] who predicted a small band gap using PEB0, which is probably caused by the structural relaxation and the availability of many allotropes of borophene nanostructures. Hybrid functional calculations in quantum espresso take much longer computational time (~50-100 times) than traditional local density and general gradient approximations as the exchange-correlation functional, and thus it requires significantly more supercomputer resources if one wants to perform hybrid functional DFT calculations using quantum espresso. We expect in the near future to use this rigorous multi-hybrid functional approach to assess which borophene allotropes may provide significant and reproducible bandgaps.

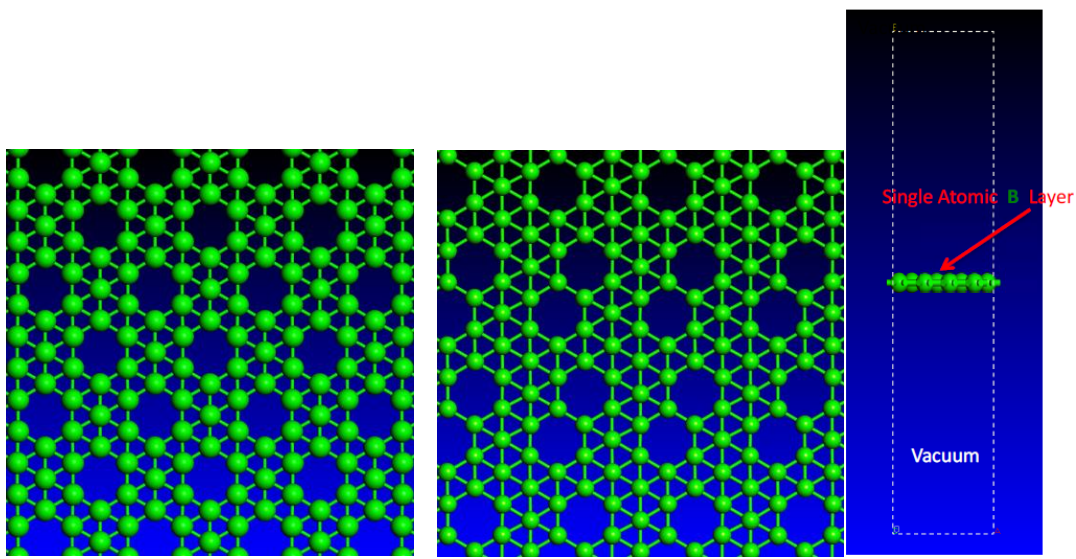

Figure 1: Top views of $\alpha$–1 borophene (left diagram) and $\beta$–1 borophene (middle diagram) atomic structures. Boron atoms are green. Side view of the 2D atomic layer of borophene (left diagram) computational structure used in DFT calculations.

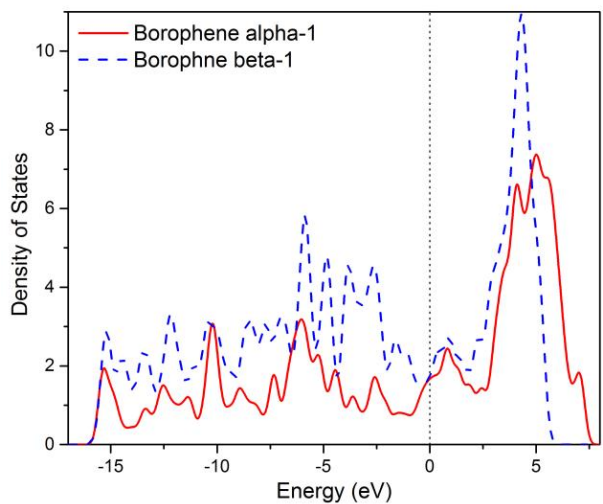

Figure 2: Calculated density of states (DOS) using density functional theory (DFT) for two borophene allotropes $\alpha$-1 and $\beta$-1. The peaks above and below the Fermi level $E_F$ at 0 eV suggest conduction and valence band action as in a semiconductor. Non-zero DOS at $\varepsilon_F$ indicates an imperfect bandgap. Some allotropes have zero here. Used the PBE functional.

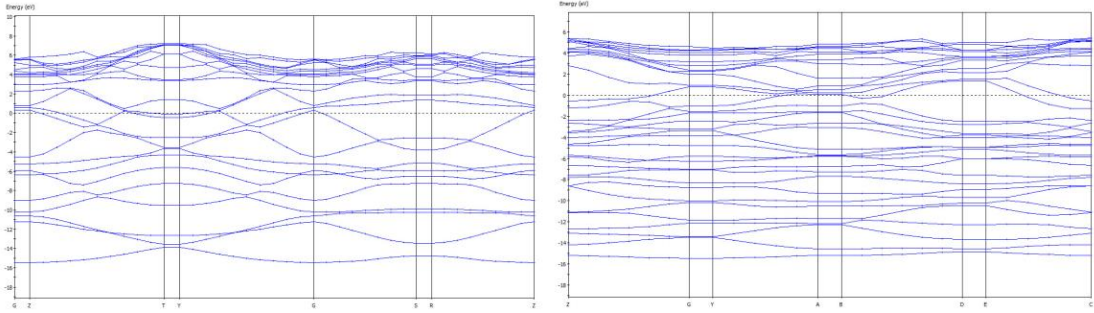

Figure 3: Calculated energy $\varepsilon(\mathbf{k})$ versus $\mathbf{k}$ (through G, Z, T, Y, G, S, R, Z symmetry points in Brillouin Zone) for $\alpha$–1 borophene (left plot) and $\beta$–1 borophene (right plot) allotropes. Used Perdew-Burke-Ernzerhof (PBE) functional which belongs to a generalized gradient approximation (GGA) for exchange correlation energy.

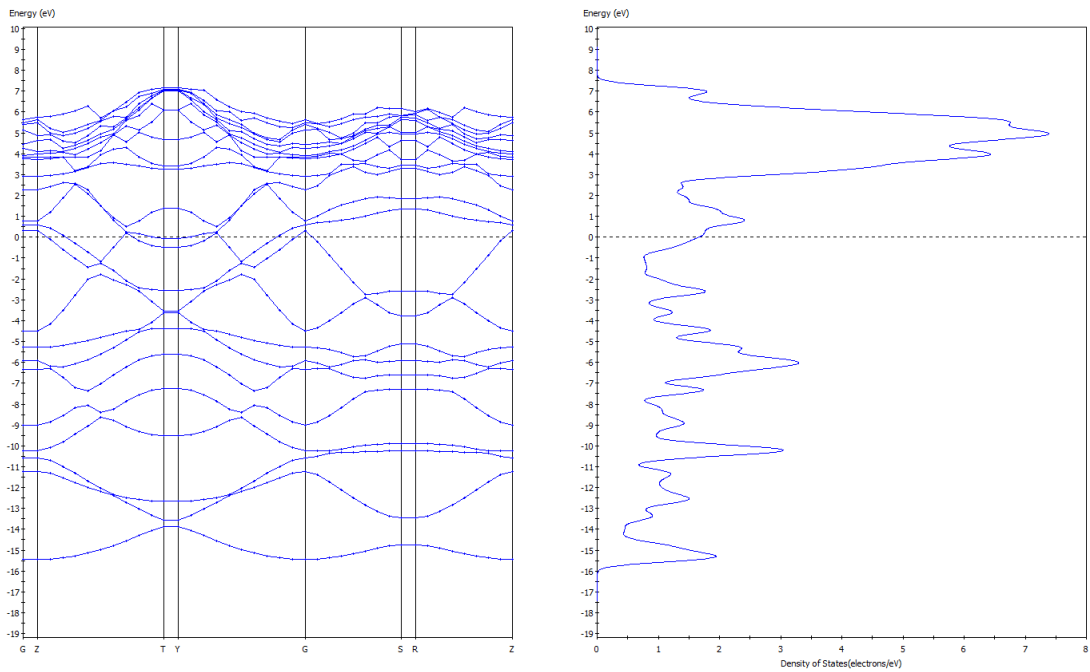

Figure 4: Calculated band structure (left) and density of states (right) of alpha-1 nanostructure using PBE GGA as the exchange-correlation functional.

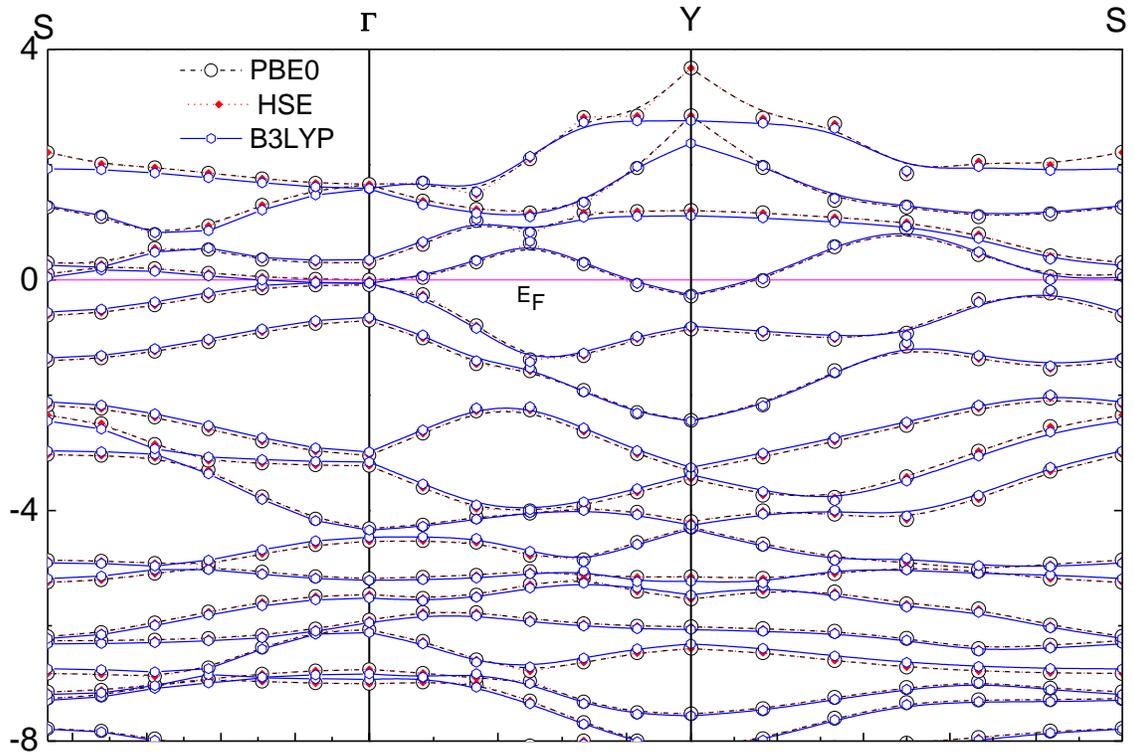

Fig. 5: Detailed comparison of the calculated band structures of alpha-1 borophene nanostructure using PEB0, HSE, and B3LYP hybrid functional DFT calculations.